\documentclass[aps,prb,twocolumn,groupedaddress,showpacs]{revtex4}

\usepackage{graphicx} 
\graphicspath{{FiguresLVDT/}}

\begin{document}


\title{Deformation of Silica Aerogel During Fluid Adsorption}


\author{Tobias Herman}
\email[]{therman@phys.ualberta.ca}
\author{James Day}
\author{John Beamish}
\affiliation{Department of Physics, University of Alberta,
Edmonton, Alberta, Canada, T6G 2J1}


\date{\today}

\begin{abstract}

Aerogels are very compliant materials --- even small stresses can
lead to large deformations.  In this paper we present measurements
of the linear deformation of high porosity aerogels during
adsorption of low surface tension fluids, performed using a Linear
Variable Differential Transformer (LVDT).  We show that the degree
of deformation of the aerogel during capillary condensation scales
with the surface tension, and extract the bulk modulus of the gel
from the data. Furthermore we suggest limits on safe temperatures
for filling and emptying low density aerogels with helium.

\end{abstract}


\maketitle

\section{Introduction}

Porous media have large surface areas, and consequently
interfacial energy contributes significantly to their behavior.
The energetic cost of the solid-vacuum interface in an empty
porous medium produces a stress on the matrix; when fluid is
adsorbed the interfacial energy may decrease, reducing the stress
on the matrix and allowing it to expand.  As more fluid is
adsorbed, liquid begins to capillary condense, creating a large
number of curved liquid-vapor interfaces throughout the sample.
The energy of these liquid-vapor interfaces creates another stress
which may result in contraction of the porous medium.  The
expansion of a porous medium upon adsorption of fluids has been
observed in a number of systems over the past
century\cite{Yates54,Beaume1980,Takamori82,Scherer86,Thibault95};
contraction has also been observed\cite{Dolino96}.

Aerogels present a somewhat different system than the denser
porous media since they are composed of a very low density network
of silica strands, often with total porosities over 95\%.  As such
they have very low elastic constants, with most elastic strain
being accounted for by the bending and twisting of the
strands\cite{Ma2002} rather than compression of the silica making
up those strands.  Therefore the effective elastic constants for
aerogels can be orders of magnitude smaller than bulk silica, and
aerogels are very sensitive to small changes in stress caused by
interfacial energy.

A porous medium exposed to vapor at low pressure collects a thin
film on the surface of the pores. As vapor pressure is increased
this film thickens slowly until the fluid suddenly capillary
condenses at some pressure below bulk saturation. Liquid invades
the pores over a narrow pressure range, driven by the pressure
difference across the curved liquid-vapor interface within the
pores. While this pressure difference may be small compared to the
elastic constants of typical solids, it can be comparable to the
bulk moduli of low density aerogels. In such a situation the
forces generated by the liquid-vapor interface during capillary
condensation can deform the medium measurably.  The capillary
forces generated by water in aerogel are sufficient to completely
crush the gel --- a fact known to anyone who has accidentally
gotten a drop of water on a hydrophilic aerogel.

Deformation of aerogel during liquid nitrogen adsorption and
desorption has been measured by Reichenauer and
Scherer\cite{Reichenauer01,Reichenauer00} for some aerogels with
densities between $150-240\frac{kg}{m^3}$ (porosities between 88\%
and 93\%) with a view to incorporating the distortion of the
aerogel into existent methods of determining pore sizes from N$_2$
adsorption isotherms. The aerogels in that study had Young's
moduli of $3.8$MPa (implying a bulk modulus of about 2MPa) and
greater, much larger than the aerogels in this study. Their
aerogels exhibited large changes in volume during capillary
condensation, and were permanently damaged by the process.  They
also observed that information about the elastic properties of
their samples could be extracted from the isotherms.

The high surface tension and large contact angle of mercury
prevents it from entering the pores of aerogels when placed in a
mercury porosimeter.  In this case the surface tension is so great
that when the aerogel is subjected to large pressures it
plastically deforms\cite{Beurroies98,Woignier97}, down to a
fraction of its original volume.

Shen and Monson performed a Monte Carlo study of fluid adsorption
in a flexible porous network which resembled a high porosity
aerogel\cite{Shen02}. Their simulated adsorption isotherms showed
that the flexibility of the network had a large effect on the
adsorption isotherms, and that the network exhibited a large
volumetric change, especially during desorption.

Aerogels are widely used as a method to introduce a ``quenched
impurity'' into a fluid, often with the goal of exploring the
effect of fixed disorder on fluid order and phase transitions.
Experiments with quantum fluids have included
$^3$He\cite{Matsumoto97,Porto99}, $^4$He\cite{Chan88,Yoon98}, and
$^3$He-$^4$He mixtures\cite{Chan96,Kim93,Mulders95}.  Liquid
crystals in aerogels\cite{Bellini01} have also been widely
studied.  Implicit in all of these studies is an assumption that
the structure of the aerogel is not affected by the fluid within
its pores, nor is the aerogel altered during filling or emptying.
While studying the temperature and porosity dependence of
adsorption isotherms in aerogels\cite{Herman05a}, we noticed that
the isotherm shapes might be showing effects from the deformation
of the aerogel by the adsorbed fluid.  This study is aimed at
quantifying the degree of aerogel deformation during helium
adsorption at a variety of temperatures.

We have measured the macroscopic linear strain in two different
low density aerogels during adsorption and desorption of low
surface tension fluids. Our measurements included an isotherm of
neon adsorbed in a $51 \frac{kg}{m^3}$ ($\sim 98\%$ porosity)
silica aerogel at 43K and several isotherms of helium adsorbed in
a $110 \frac{kg}{m^3}$ (95\% porosity) silica aerogel at
temperatures from 2.4K to 5.0K. At these temperatures the fluid
surface tensions were small, but the low elastic constants of the
aerogels allowed a significant deformation nevertheless.  The
compression of the aerogel was greatest during desorption.  From
the isotherms we show that the degree of compression during
capillary condensation within the aerogel is directly related to
the surface tension of the adsorbate. Furthermore, the bulk
modulus of the aerogel was extracted from the adsorption and
desorption isotherms.

The compression due to capillary condensation has not previously
been studied in such compliant materials.  By using high porosity
aerogels, we were able to directly observe, for the first time,
the deformation associated with low surface tension fluids like
helium.  By making measurements very slowly, we avoided rate
dependent effects and were able to study the hysteresis between
compression during filling and emptying of aerogels.  Our
measurements extended close to helium's critical point, which
allowed us to vary the surface tension over a wide range compared
to previous measurements, which used nitrogen and much more rigid
samples.  From our results, it is clear that the deformation
associated with surface tension must be taken into account when
interpreting adsorption isotherms and to avoid damage when filling
high porosity aerogels.

\section{Experimental Method}
The aerogel samples were synthesized in our lab from tetramethyl
orthosilicate (TMOS) using a standard one-step base catalyzed
method followed by supercritical extraction of the methanol
solvent\cite{Poelz82}. The two aerogels studied had densities of
$110 \frac{kg}{m^3}$ and $51 \frac{kg}{m^3}$, corresponding to
porosities of 95\% and slightly less than 98\% respectively.  They
are referred to as aerogels ``B110'' and ``B51'' throughout this
paper.  Both aerogel samples were cut into cylinders about 1cm
long. B51, used in the first experiment, was about 1.2cm in
diameter --- it equilibrated so slowly that it took weeks to
measure a single isotherm. To reach thermal equilibrium more
quickly, a smaller sample of B110 (4mm in diameter) was used for
the second experiment.

Our linear variable differential transformer (LVDT) allowed us to
make high precision, non-contact measurements of the relative
position of a cylindrical ferromagnetic core and a set of primary
and secondary coils\footnote{The neon data were collected using
hand wound coils; the helium data were collected using a Schevitz
HR050 sensor, available from Measurement Specialties, Inc..  The
LVDT output was measured using an LR400 mutual inductance bridge,
from Linear Research Inc.}.  The response of the LVDT was highly
temperature dependent, and room temperature calibrations were not
used to interpret the low temperature data. In the liquid helium
cryostat it was possible to make a direct calibration of the LVDT,
but for measurements with neon an approximate calibration had to
be computed from the adsorption isotherm itself (as will be
explained later).

Two experimental cells were used over the course of this
experiment to accommodate the two different aerogel samples. Both
cells were made of copper and had the same general layout, with
the LVDT core supported about 2cm above the sample by a thin brass
rod.  The support was kept in contact with the sample by gravity,
and the cells kept upright once assembled.  The initial position
of the external LVDT coils relative to the core was controlled
through set screws.

In our initial experiment neon was admitted to the experimental
cell containing B51 in volumetric shots from a room temperature
gas handling system, and pressure measurements were made with a
room temperature gauge\footnote{Mensor 1000psi gauge, Model 4040}.
Cooling was provided by a Gifford-McMahon closed cycle
refrigerator; the temperature was controlled to $\pm 1$mK using a
platinum resistance thermometer and a thick-film heater mounted
directly on the cell.

In the second set of measurements, on helium in B110, the fluid
condensation into the system was controlled by stepping the
pressure in the cell through the use of a low temperature pressure
regulation ballast\cite{Herman05a}. This second method did not
provide information on the absolute quantity of fluid adsorbed by
the aerogel but ensured long term pressure stability. A
Straty-Adams\cite{Straty69} type capacitive pressure gauge was
used for low temperature \emph{in situ} pressure measurement. The
cell was mounted on a liquid helium cryostat; the temperature was
controlled to $\pm 50 \mu$K using a germanium resistive
thermometer and thick-film heater mounted directly on the cell.

\section{Neon at 43K in B51}
\begin{figure}
\includegraphics[width=\linewidth]{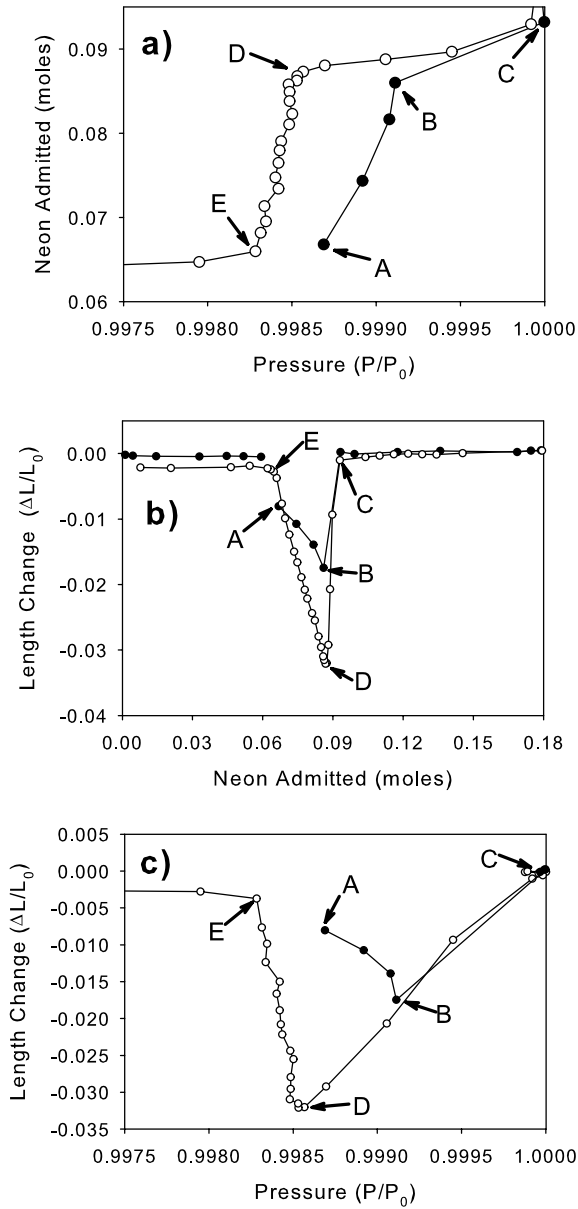}
\caption{\label{LVDTNe43K3Panes}(a)Adsorption isotherm for neon in
B51.  (b)Contrac\-tion of the aerogel as a function of neon added
to cell. (c)Contraction of the aerogel as a function of system
pressure. Data were taken during filling (solid symbols) and
emptying (open symbols). Points ``A'' and ``B'' refer to the
beginning and end of capillary condensation during filling. At
point ``C'' the gel is full of liquid but completely relaxed.  At
point ``D'' the gel is still full of liquid but the aerogel has
been compressed by the removal of some of the neon, so that its
pore volume has been reduced. At point ``E'' the aerogel is empty
of liquid, but shows permanent deformation.}
\end{figure}

The adsorption and desorption isotherms for neon at 43K in B51 are
shown in Fig.~\ref{LVDTNe43K3Panes}a.
Note that neon at 43K is very close to its critical point (T$_c
\sim 44.5$K, P$_c \sim 2.7$MPa, $\rho_c \sim 485 \frac{kg}{m^3}$).
As such, it has a very low surface tension (at 43K neon has the
same surface tension as liquid helium does at 3.7K) and its liquid
and vapor  densities differ by only a factor of three ($\rho_l =
740 \frac{kg}{m^3}$ and $\rho_v = 250 \frac{kg}{m^3}$). We could
not perform a direct calibration of the LVDT at 43K because there
was no direct access to the LVDT within the cryostat; however, as
shown by Reichenauer and
Scherer\cite{Reichenauer01,Reichenauer00,Reichenauer04},
information can be extracted about the sample shrinkage from the
adsorption isotherm itself.

Since capillary condensation occurs over a narrow pressure range
($0.998 < \frac{P}{P_0} < 1$), the liquid can be treated as
incompressible during capillary condensation. The high density
corners of the adsorption and desorption isotherms (points``B''
and ``D'' in Fig.~\ref{LVDTNe43K3Panes}a respectively) correspond
to the aerogel being full of liquid, but compressed because of
capillary forces at the surface pores. After capillary
condensation was complete in this sample, the aerogel continued to
adsorb fluid up until bulk saturation was reached (just below
point ``C''). The significant slope of the isotherm between
capillary condensation and bulk condensation (i.e.\ $0.999 <
\frac{P}{P_0} < 1$) indicates that the aerogel was still adsorbing
liquid even though there was no longer any vapor within the pores.
By measuring how much fluid was adsorbed after the completion of
capillary condensation it is possible to calculate how much
swelling occurred in the aerogel over this pressure range.  The
degree of swelling can then be used to form a calibration for the
LVDT.

The bulk neon vapor at 43K has a high density and the cell
included a large bulk volume, so the total amount of neon admitted
into the cell does not correspond to the amount of neon adsorbed
by the aerogel.  However, since we are dealing with such a small
pressure range during capillary condensation we can assume that
all neon admitted over the pressure range  $0.999 < \frac{P}{P_0}
< 1$ is adsorbed by the aerogel -- the bulk vapor density changes
very little over this range.

Assuming the deformation of the aerogel is isotropic, the change
in aerogel volume ($\Delta V$) can be related to the change in its
length ($\Delta L$) and the quantity of fluid ($\Delta n$)
adsorbed during swelling (or shrinking) by:
\begin{equation} \Delta V = \frac{\Delta n}{\rho_l - \rho_v}
= V_0 \left[ 1 -\left( 1-\frac{\Delta L}{L_0} \right)^3 \right]
\label{EqNeonCalibration}
\end{equation} where ``$L_0$'' and ``$V_0$'' are the initial
length and volume of the aerogel sample and ``$\rho_l$'' and
``$\rho_v$'' are the density of liquid and gaseous neon
respectively.  While this technique is not very precise, it does
give an estimate of the maximum linear aerogel compression which
can then be used as an LVDT calibration.

Using this calibration the aerogel length change has been plotted
as a function of the amount of neon adsorbed in
Fig.~\ref{LVDTNe43K3Panes}b and as a function of pressure in
Fig.~\ref{LVDTNe43K3Panes}c. Zero has been set to be the fully
relaxed aerogel filled with liquid.  Within the resolution of the
LVDT no deformation was seen for the first 0.06 moles of neon
admitted to the cell --- in this regime, a thin film was
collecting on the aerogel strands in equilibrium with bulk vapor.
Then, as neon began to capillary condense near point A, the
aerogel contracted reaching a maximum linear compression of almost
2\% at point B.  At this point the gel was full of liquid, but
compressed from its original volume. As more neon was added to the
cell it was adsorbed by the aerogel, allowing the gel to relax and
expand.  Once the gel was full and bulk liquid began collecting
(at point C, $n_{neon} \sim 0.095$moles), there were no longer any
curved liquid-vapor interfaces causing stress within the aerogel
and it had re-expanded to its original size. Upon desorption of
the neon, the process occurred in reverse; however during
desorption the aerogel was linearly compressed by almost 3.5\% at
point D.  By point E the aerogel had re-expanded --- in this
pressure regime only a film of neon remains on the silica strands.
However, the gel did not fully return to its original size; there
was a slight permanent deformation (about 0.2\% linear
compression).

Hysteresis is generally observed for adsorption of fluids in
porous media --- the adsorption and desorption of the fluid taking
place at different partial pressures.  The precise mechanism of
this hysteresis is an open question for aerogels, but it may
indicate a different liquid-vapor interface shape for adsorption
and desorption.  The lower partial pressure for desorption means
that there exists a greater pressure difference across the curved
liquid-vapor interface, and results in a greater maximum stress on
the aerogel during desorption.

The portion of the isotherms closest to $P_0$ are identical for
the emptying and filling branches of the isotherm; in this region
the gel is full of liquid and the aerogel matrix is swelling or
shrinking in response to the pressure difference across the
liquid-vapor interface present at the aerogel surface rather than
interfaces within the pores.

\section{Helium in Aerogel B110}

The second experimental cell, used to investigate helium
adsorption in aerogel, allowed more precise compression and
pressure measurements. It also allowed a direct low temperature
calibration for the LVDT.  The LVDT calibration was reproducible
to within about 1\% upon thermal cycling. Since the sample had a
smaller diameter than the B51 sample, the pressure exerted by the
weight of the ferromagnetic core was larger.  To avoid overloading
the sample we used a denser aerogel, B110, in this cell; the gel
was compressed about 0.3\% by the weight of the core ($\sim 3$g).

\begin{figure}
\includegraphics[width=\linewidth]{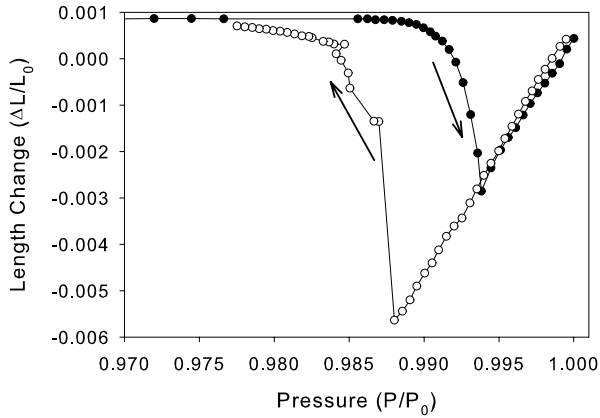}
\caption{\label{LVDTHe4K}Deformation of B110 during adsorption
(solid symbols) and desorption (open symbols) of helium at 4.2K.}
\end{figure}

An isotherm for helium in B110 at 4.200K is shown as
Fig.~\ref{LVDTHe4K}.  
Since the pressure in the cell is controlled rather than the
quantity of helium,  the isotherm is only plotted with length
change as a function of cell pressure, as in
Fig.~\ref{LVDTNe43K3Panes}c.  During the low pressure, $P/P_0
<0.95$, formation of a thin film of helium the aerogel expanded by
about 0.08\%. Equilibration took many hours for these films, so
equilibrium data of the initial dilation of aerogel upon helium
adsorption were not collected.  Similar long equilibration times
for helium films in aerogels have been observed before, as has the
dilation\cite{Thibault95}.  Equilibration of points within the
hysteretic region was also very slow\cite{Herman05a}.  The maximum
contraction occurred during desorption and was equal to about
0.6\% of total length.

\begin{figure}
\includegraphics[width=\linewidth]{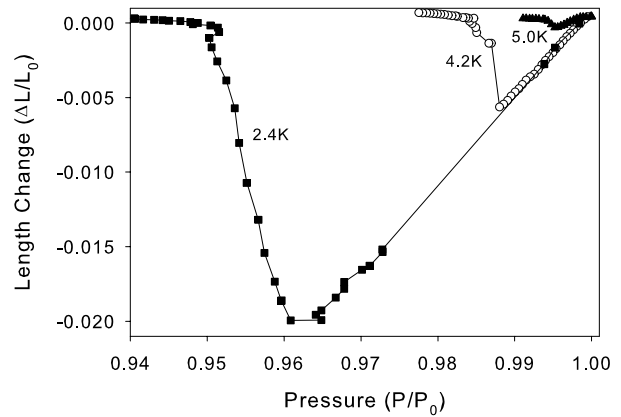}
\caption{\label{LVDTEmptying} Deformation of B110 during emptying
isotherms at 2.4K, 4.2K, and 5.0K.}
\end{figure}

Figure~\ref{LVDTEmptying} 
shows three desorption isotherms taken at 2.4K, 4.2K and 5.0K
respectively.  As the temperature was raised, the surface tension
of the helium decreased, causing less deformation of the aerogel.
The maximum deformation decreased by more than a factor of twenty
between 2.4K and 5.0K.  No permanent compression was observed for
this sample.  Above the critical temperature of helium
(T$_c$=5.195K) no contraction was observed, as would be expected
without a liquid-vapor interface.

\section{Analysis/Discussion}

\subsection{Dilation at low vapor pressure}
Consistent with an earlier study\cite{Thibault95}, dilation of the
aerogel during the initial stages of helium adsorption was
observed at all temperatures studied.  The maximum dilation
depended much less on temperature than did the deformation during
capillary condensation
--- in all cases it was just under 0.1\% of the total sample
length. This is consistent with the dilation being governed by the
energetics of adsorption sites on the silica strands, which can
involve binding energies much larger the liquid-vapor interfacial
energy or the thermal energy at liquid helium temperatures.

\subsection{Aerogel bulk modulus}
During fluid adsorption and desorption, curved liquid-vapor
interfaces abound within the open aerogel pore structure.  A
Laplace pressure exists across each of these interfaces, which can
place stress on the aerogel.  The Laplace pressure is a purely
mechanical construct, with the surface tension ($\sigma_{lv}$)
across the interface responsible for the pressure difference
between the liquid ($P_l$) and vapor ($P_v$) phases:
\begin{equation} \Delta P \ = \ P_l - P_v \ = \ \frac{2
\sigma_{lv}}{r} \end{equation} assuming a hemispherical interface
or radius $r$. Usually the porous medium is assumed to be rigid,
but aerogels are extraordinarily compliant, and any stresses must
be balanced by deformation of the aerogel.    Once the aerogel is
completely filled with liquid, the liquid-vapor interface exists
only at the aerogel surface and the Laplace pressure is felt as a
macroscopic stress.

The Laplace pressure also results in undersaturation --- the
condensation of fluids below bulk saturation pressure ($P_0$).
This behavior is described by the Kelvin equation, which can be
expressed in many forms. The simplest form for our
purposes\cite{LandauStatPhys} relates the stress caused by the
fluid interface to the vapor pressure in the cell:
\begin{equation} P_l - P_v = \left(
\frac{V_v-V_l}{V_l} \right) \left( P_v-P_0 \right)
\label{EqKelvin2}
\end{equation}
Here $P_l$ and $P_v$ are the pressures, and $V_l$ and $V_v$ the
molar volumes, of the liquid and vapor phases respectively.  This
derivation assumes that the liquid and vapor are incompressible,
which is a fairly accurate approximation given the small pressure
ranges over which capillary condensation occurs in aerogels.

When the Laplace pressure acts as a macroscopic stress, then the
Kelvin equation can be used to relate the stress on the aerogel to
the vapor pressure in the cell.  Thus Eq.~\ref{EqKelvin2} can be
combined with the measured slope of the isotherms after capillary
condensation has occurred (e.g.\ for $0.995 < \frac{P}{P_0} < 1$
for helium in B110 at 4.2K) to calculate the elastic properties of
the aerogel. In this regime the gel was full of liquid and the
pressure difference between the liquid and vapor phases was
balanced by the elastic stress of the aerogel. If the deformation
of the aerogels is isotropic then the bulk modulus, K$_{gel}$, can
be used to characterize its response to the stress exerted by the
liquid-vapor interface.

For high porosity aerogels and small deformations (i.e.\
$\frac{\Delta L}{L_0} \approx \frac{\Delta V}{3 V_0}  $) the
sample experiences this pressure difference as a compressive
stress, which gives:
\begin{equation} \frac{\Delta L}{L_0}   \approx \frac{1}{3}
\frac{1}{K_{gel}} \left( \frac{V_v-V_l}{ V_l} \right) \left(
P_v-P_0 \right) \label{EqBulkMod}
\end{equation} Thus we can extract a value for K$_{gel}$
from the slope of each isotherm when plotted as $\Delta L / L_0 \
vs \ P_v$, without needing to know either the fluid's surface
tension or the aerogel's pore size.  In fact, measurements on this
portion of our isotherm, where the aerogel is full but partially
compressed, cannot tell us about the effective pore size.  That
information (R$_{cap}$) comes from the ``breakthrough radius,''
the point at which the meniscus curvature $r$ becomes equal to an
effective pore size during desorption and the interface becomes
unstable so that the pores suddenly empty.

If we know the liquid's surface tension and the pressure, $P_v$,
at which pores begin to empty, we can determine a pore size,
$R_{cap}$. In fact, this is a standard way of determining pore
size from isotherms --- there is no need to measure sample
deformation $\Delta$L, the desorption pressure is all that is
necessary. Assuming the simplest (hemispherical) form for the
meniscus, then at breakthrough: \begin{equation} \left( P_0 - P_v
\right) = \left( \frac{V_l}{V_v -V_l} \right) \frac{2
\sigma_{lv}}{R_{cap}}\label{Eq:breakthrough}
\end{equation}
A similar calculation can define a radius during adsorption,
although its meaning is less clear.

This derivation assumes that the liquid and vapor are
incompressible, which is a fairly accurate approximation given the
small pressure ranges over which capillary condensation occurs in
aerogels.  More commonly one assumes the vapor behaves like an
ideal gas and the liquid molar volume is much smaller and can be
neglected.  This gives the more familiar form of the Kelvin
equation:
\begin{equation}
-RT\ln{\frac{P_v}{P_0}} = -V_l\, \frac{2 \sigma_{lv}}{r} + V_l
\left( P_0-P_v \right)\label{EqKelvin}
\end{equation}
Neither of these assumptions is appropriate near the liquid-vapor
critical point, so we use the form in Eq.~\ref{Eq:breakthrough}
rather than Eq.~\ref{EqKelvin}.

What our aerogel pore size analysis (from the ``breakthrough''
pressure, $P_v$) does is show that this rather macroscopic,
classical treatment is valid over a wide range of $\sigma_{lv}$
near T$_c$ (i.e.\ the value for $R_{cap}$ is essentially the same,
even though $\sigma_{lv}$ varies by nearly twenty times).  This
also shows that, despite their unique tenuous structure, quite
unlike an array of uniform pores, a description of capillary
condensation in terms of a single effective pore size seems
adequate.

\begin{figure}
\includegraphics[width=\linewidth]{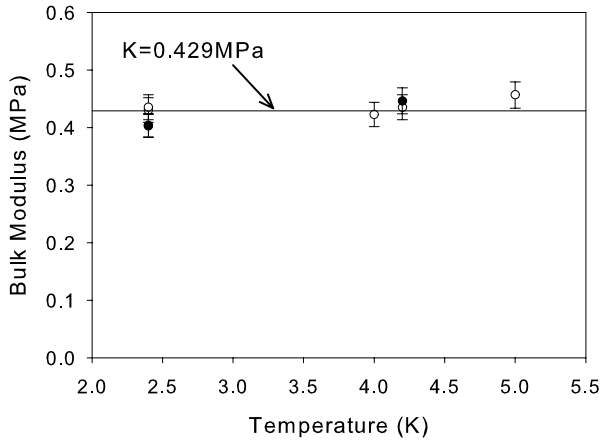}
\caption{\label{LVDTModulus} The bulk modulus of aerogel B110 as
calculated from adsorption (solid symbols) and desorption (open
symbols) isotherms using Eq.\ref{EqBulkMod}.}
\end{figure}

The bulk moduli calculated from several helium adsorption
isotherms in B110 are plotted in Fig.~\ref{LVDTModulus}.
 Calculations were made using all adsorption and desorption
isotherms separately over a temperature range where $\sigma_{lv}$
changes by a factor of eighteen, confirming the validity of
Eq.~\ref{EqBulkMod}. The values extracted for K$_{gel}$ are
roughly constant, and the mean value (K$_{gel} \approx 0.43$MPa)
has been included on the plot as a solid line.

The bulk modulus for silica aerogel depends sensitively on aerogel
density, and can be extracted from measurements of the Young's
modulus (E)~\cite{Woignier89} or shear modulus
(G)~\cite{Daughton03} using: \begin{equation}  K =
\frac{E}{3(1-2\nu)} = \frac{2(1-\nu)G}{3(1-2\nu)} \end{equation}
These equations require knowledge of the Poisson's ratio ($\nu$)
for aerogel, which is about 0.2~\cite{Gross88}. Pure silica has a
bulk modulus of roughly $3.5 \times 10^4$MPa, much larger than
aerogel. The moduli for base catalyzed silica aerogels with
densities used in this study ($51 \frac{kg}{m^3}$ and $110
\frac{kg}{m^3}$) should be about 0.08MPa and 0.8MPa respectively,
although these values can vary greatly between samples. Our value
for the bulk modulus of B110 of 0.43MPa is consistent with these
values, although it is a little lower than expected.  A similar
calculation for the data in Fig.~\ref{LVDTNe43K3Panes} yields a
bulk modulus of K=0.065MPa for aerogel B51.

\subsection{Maximum deformation}
\begin{figure}
\includegraphics[width=\linewidth]{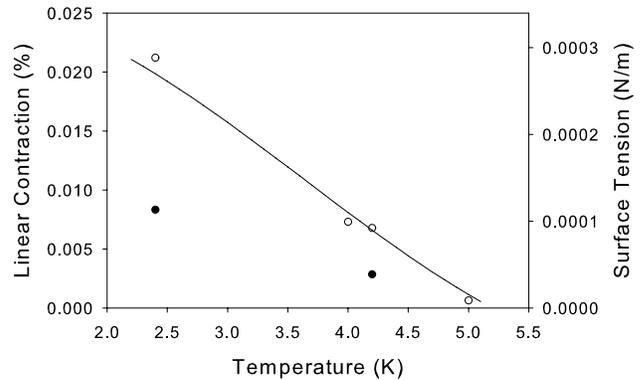}
\caption{\label{LVDTTotalDef} The amount that sample contracted
during adsorption (solid symbols) and desorption (open symbols) is
shown here as a function of temperature.  The temperature
dependence scales roughly with helium surface tension (which is
nearly linear in this temperature range), shown as a solid line
for comparison.}
\end{figure}

Since helium adsorption isotherms in B110 were collected at
several temperatures, the temperature dependence of the maximum
compression during capillary condensation could be analyzed;
Fig.~\ref{LVDTTotalDef}
shows the degree of compression during capillary condensation, as
well as the bulk surface tension at each temperature. While the
degree of contraction depends sensitively on temperature, it
scales roughly with surface tension (plotted as a solid line in
Fig.~\ref{LVDTTotalDef}.

The scaling of maximum deformation with surface tension is
consistent with a characteristic ``breakthrough radius''
describing the curvature of the liquid-vapor interface at the
surface of the aerogel just prior to the percolation of the vapor
phase into the sample (invasion of the helium vapor phase into
aerogel during desorption has been observed\cite{Lambert04} in
optical experiments). While it is not clear how this breakthrough
radius relates to the aerogel structure, the magnitude of this
breakthrough radius can be estimated from the desorption isotherm
using the Kelvin Equation. A similar calculation can be performed
using the pressure at which capillary condensation is complete
along the adsorption isotherms, although the form of the
liquid-vapor interface during filling is even less clear.
Table~\ref{TableRadii} summarizes the values of the interface
radii calculated from our isotherms, as well as the effective
capillary pressure and predicted compression of the aerogels.
\begin{table*}
\begin{tabular}{|l|c|ccc|} \hline
 & \textbf{B51 -- Neon} && \textbf{B110 -- Helium}& \\
 & (43K)  &   (2.4K)&   (4.2K) &   (5.0K) \\
\hline
Breakthrough radius, Desorption (nm) &  40(4)   & 21(1)  &  22(1) & 20(1)    \\
``Breakthrough radius,'' Adsorption (nm) &  68(7)   & N/A  &
44(4) & N/A    \\
\hline
Surface tension (\textbf{$\sigma_{lv}) \left( \frac{mN}{m} \right)$}   &  0.13  &  0.27  &  0.090  &  0.015   \\
\hline
Capillary pressure ($P_{cap} = 2\sigma_{lv}/R_{Des}$) (MPa) & 0.0064 & 0.026 & 0.0082 & 0.0015\\
Calculated Bulk Modulus ($K_{gel}$) (MPa) & 0.065 &
0.43 &0.43 & 0.43  \\
 Measured volumetric  compression  &
9\% & 6\% & 2\% & 0.2\%  \\\hline
\end{tabular}
\caption{\label{TableRadii}Liquid-vapor interface radii at
completion of capillary condensation (or beginning of capillary
evaporation) calculated using the Kelvin equation, for adsorption
and desorption isotherms shown in this paper.  Errors in the last
digit are given in parentheses.  Also the effective capillary
pressures acting on the samples at breakthrough, aerogel bulk
moduli, and maximum volumetric contraction during desorption.}
\end{table*}

The capillary pressure generated by the curved liquid-vapor
interface during desorption is significant compared to the bulk
moduli of the aerogels. It is easy to see how higher surface
tension fluids can easily damage aerogels --- assuming a
breakthrough radius of 20nm, water (295K) and liquid nitrogen
(77K) would generate capillary pressures of 7MPa and 0.9MPa
respectively, much larger than the bulk moduli of these aerogels.

\subsection{Damage to aerogels}

No permanent densification was seen in our denser aerogel (B110)
over the range of this experiment.  The maximum volumetric
compression at any time was about 6\%, which is generally within
the elastic regime for aerogels. However, B51 appeared to exhibit
damage after desorption of neon. It had experienced a 10\%
volumetric compression, enough to cause permanent structural
changes in the aerogel.  The manner in which the maximum
compression scales with surface tension allows us to predict the
compression for a given aerogel, fluid, and temperature.  This is
especially important for very low density gels, with their
associated low bulk moduli --- for such samples permanent damage
could result from the surface tension of liquid helium.  At common
filling temperatures, such as 4.2K, gels with porosities or 98\%
or more will likely be damaged.

Compression of aerogels at room temperature leads to bulk
densification and important microstructural changes in the sample.
Such changes have been investigated\cite{Beurroies98,Woignier97}
during the room temperature densification of silica aerogels in a
mercury porosimeter.  Three characteristics of the aerogel
structure were measured --- the fractal dimension, the solid
particle size, and the cluster size (or correlation length).  As
the aerogel was compressed the particle size remained unchanged,
consistent with the silica particles (that compose the aerogel
strands) being relatively unaffected by small pressures.  However,
the fractal dimension of the aerogels increased slightly, and the
correlation length decreased significantly, during compression.
These three factors are all very important in determining how
fluids are affected by the presence of aerogel, and they are
usually assumed to remain constant throughout an experiment.
However, if the aerogel is compressed during the experiment these
factors may change.

The increase in aerogel density is generally assumed to be due to
the preferential collapse of the largest ``pores,'' with
relatively little damage to the smallest scale structures.
However, the presence of large, open, pores is what distinguishes
aerogel from other porous media and any damage to this structure
may have dramatic effects on the behavior of fluids within the
aerogels. To avoid such damage, one must be careful to avoid
capillary stresses that could deform the gel beyond its elastic
limit.  The safest way to fill or empty an aerogel is above the
liquid-vapor critical point, where no liquid-vapor interface
exists.  However, using calculations such as those shown in
Table~\ref{TableRadii} allows one to estimate the degree of
deformation induced during filling and emptying at lower
temperatures, and a filling temperature can be chosen that does
not allow the sample to experience plastic deformation.

\subsection{Shape of adsorption isotherms}
Note that the amount of neon adsorbed or desorbed at 43K during
capillary condensation --- the steepest portion of the curve at
$P/P_0 \sim 0.9990$(adsorption) or $0.9983$(desorption) --- is
about 65\% of the total neon within the gel.  The volumetric
compression of the aerogel by about 10\% during desorption is a
significant portion of this, so that compression is a very
important part of the capillary condensation behavior.
Qualitatively the effect of aerogel compliance is to make the
capillary condensation/desorption portions smaller (i.e.\ the
fluid density change from point A to point B in
Fig.~\ref{LVDTNe43K3Panes} would be spread out over a wider range,
from point A to point C if the aerogel was rigid). Simultaneously,
the isotherms become steeper in a compliant material like aerogels
since effective pore radius decreases during compression (by up to
3.5\%).

\section{Summary}
We have investigated the effect of small capillary pressures on
the deformation of low density silica aerogels during adsorption
and desorption of fluids.  For most conditions in this study, this
deformation was found to be completely elastic, but in the lower
density aerogel (B51) permanent damage was seen.  The compression
of the aerogel during adsorption and desorption can be used to
compute the bulk modulus of the aerogel, with no knowledge of the
aerogel pore structure being necessary.

The highly compliant nature of low density aerogels also has
important implications for the shape of their adsorption
isotherms.   This effect becomes less important as the degree of
compression is reduced --- it is not very significant in the
helium adsorption/desorption isotherms in B110.  This effect may
be further reduced if the aerogel is not free to deform, such as
when grown within the small pores of a metal sinter.

Finally, it should be pointed out that when low density silica
aerogels are used as a method to introduce disorder into fluid
systems such as helium or liquid crystals the aerogels may be
damaged by any fluid interfaces present. Even the low surface
tensions of $^3$He and $^4$He are capable of damaging low density
aerogels during filling and emptying.

\end{document}